\documentclass[english]{emulateapj}
\usepackage[T1]{fontenc}
\usepackage[latin9]{inputenc}
\usepackage{graphicx}
\usepackage{amssymb}

\makeatletter

\providecommand{\tabularnewline}{\\}

\makeatother

\makeatother

\usepackage{babel}

\makeatother

\usepackage{babel}

\makeatother

\usepackage{babel}

\begin{document}
\shorttitle{Origin of the B-stars in the Galactic center}
\shortauthors{Perets \& Gualandris}
\title{Dynamical constraints on the origin of the young B-stars in the Galactic
center}

\author{Hagai B. Perets\altaffilmark{1} and Alessia Gualandris\altaffilmark{2}}

\email{hperets@cfa.harvard.edu}

\altaffiltext{1}{CfA fellow, Harvard-Smithsonian Center for Astrophysics, 60 Garden St., Camridge, MA 02138, USA}

\altaffiltext{2}{Max-Planck Institut für Astrophysik, Karl-Schwarzschild-Str. 1, D-85741 Garching, Germany}

\begin{abstract}
Regular star formation is thought to be inhibited close to the massive
black hole (MBH) in the Galactic center. Nevertheless, tens of young
main sequence B stars have been observed in an isotropic distribution
close to it. These stars are observed to have an apparently continuous
distribution from very close to the MBH $<0.01$ pc and up to at least
$\sim0.5$ pc, suggesting a common origin. Various models have been
suggested for the formation of the B-stars closest to the MBH ($<0.05$
pc; the S-stars), typically involving the migration of these stars
from their original birthplace to their currently observed position.
Here we explore the orbital phase space distribution of the B-stars
throughout the central $pc$ expected from the various suggested models
for the origin of the B-stars. We find that most of these models have
difficulties in explaining, by themselves, both the population of
the S-stars ($<0.05$ pc), and the population of the young B-stars
further away (up to $0.5$ pc). Most models grossly over-predict the
number of B-stars up to $0.5$ pc, given the observed number of S-stars.
Such models include the intermediate-mass black hole assisted cluster
inspiral scenario, Kozai-like perturbations by two disks, spiral density
waves migration in a gaseous disk, and some of the eccentric disk
instability models. We focus on one of the other models, the massive
perturber induced binary disruption, which is consistent with both
the S-stars and the extended population of B-stars further away. For
this model we use analytical arguments and N-body simulations to provide
further observational predictions. These could be compared with future
observations to further support this model, constrain it or refute
it. These predictions include the radial distribution of the young
B-stars, their eccentricity distribution and its dependence on distance
from the MBH (higher eccentricities at larger distances from the MBH),
as well as less specific expectations regarding their mass function. 
\end{abstract}

\section{Introduction}

In recent years high resolution observations have revealed the existence
of many young O and B stars in the galactic center (GC). Accurate
measurements of the orbital parameters of these stars give strong
evidence for the existence of a massive black hole (MBH) which governs
the dynamics in the GC \citep{ghe+98,eis+05,gil+09}. Most of the
young stars are observed in the central 0.5 pc around the MBH. The
young stars population in the central $\sim$arcsecond ($0.05$ pc),
the so called 'S-stars', contain only young B-stars (likely masses
of $7-15$ $M_{\odot}$), in apparently isotropic distribution around
the MBH, with relatively high eccentricities (consistent with a thermal
distribution, with $0.3\le e\le0.95$; \citealt{gil+09}). The young
stars outside this region contain many O-stars and present markedly
different orbital properties, with a top heavy mass function and a
disk (or disks) like configuration(s) \citep{lev+03,lu+09,bar+09,bar+10},
and much more circular orbits. The sharp difference and discontinuity
between the O-stars population outside the central $0.05$ pc, and
the S-stars population inside this region strongly suggest a different
origin for these two populations (see also \citealp{bar+10}).

Recently, it was found that the isotropically distributed B-stars
population extends (apparently) continuously beyond the central arcsecond
and up to at least $0.5$ pc from the MBH \citep{bar+10}. Although
many studies have explored the origin of the 'S-stars' (see \citealt{per+09b}),
the observations of the extended population may shed new light on
the origin of these stars. Given the seemingly continuous nature of
the B-star population, it is likely that this population of both the
S-stars and the extended population at larger distances (or at least
most of it), has a single origin. In the following we make this assumption
and study possible scenarios for a \emph{single} common origin of
all the GC B-stars and their later evolution. We discuss and suggest
specific observational constraints on the various models. We then
mainly focus on one of these models, the massive perturbers induced
binary disruption \citep{per+07}, and use analytical arguments and
N-body simulations to provide specific predictions and observational
signatures expected from this model.

\section{Origins and evolution}

Regular star formation in the close region near the MBH is inhibited
by the tidal forces which destroy any potential giant molecular cloud
progenitor. For this reason, young stars in this region are thought
to either form elsewhere, further away from the MBH, and then migrate
rapidly closer to the MBH; or form in-situ near the MBH, in an irregular
form of star formation, most likely through instabilities in a gaseous
disk (see below). After their formation/capture, stars in the GC environment
can dynamically evolve. Each of the scenarios for the B-stars origin
produces a different and unique distribution of their initial orbital
configurations. These configurations, however, can later dynamically
evolve due to relaxation processes. The currently observed properties
of these stars therefore reflect the combination of the initial configurations
and their later evolution. In the following we discuss the various
models for the origin of the B-stars. We briefly review each of them,
discuss their possible caveats, and detail dynamical constraints which
could be used to exclude or support these models. As we shall see,
simple efficiency arguments show that most of the suggested models
for the B-stars origin can already be strongly constrained, if not
excluded, by current observations.

Note that although we focus on the origin of the B-stars in the GC,
in the following we also mention scenarios for the disk O and WR-stars
origin, since their population is closely related to that of the B-stars
in some of the models we discuss.

\subsection{In-situ star-formation from infalling gaseous clumps}

It was suggested that the young OB stars in the GC were formed in-situ
in the central pc a few million years ago in gaseous disks and/or
streams formed from infalling and/or colliding gaseous clumps \citep{mor93}.
Analytic calculations and simulations \citep{nay+05b,lev07,bon+08,war+08,hob+09}
have shown that stars could form in such fragmenting clumps to produce
stellar disks and/or other coherent structures, in the region of a
few 0.01 pcs up to a few 0.1 pcs from the MBH. These could, in principle,
be consistent with the observed young stellar structures (clockwise
disk and a secondary inclined structure). This scenario could also
explain the radial distribution of stars in the observed stellar disk
and possibly their eccentricities. Moreover, some of the proposed
models produce a top heavy mass function (MF) for the newly formed
stars, possibly consistent with observations \citep{bar+10}. None
of these, however, produces the population of B-stars within $0.05$
pc of the MBH.

One should note that our poor understanding of the initial conditions
and the star formation processes in the GC, allows for a wide parameter
space to choose from, which, naturally, raises difficulties in constraining
(or strictly falsifying) such models. Nevertheless, the robustness
of producing some sort of star formation in the GC region under a
variety of conditions explored in the literature suggest these models
as the currently most promising scenarios for the origin of the young
stellar disk and stellar structures (although less likely the origin
of the isotropic B-stars population, including the S-stars, as discussed
below).

\subsection{In-situ star-formation followed by rapid migration}

None of the models for in-situ star formation in the central pc suggests
the formation of stars as close to the MBH as the S-stars, or the
apparent existence of two distinct young O and B stellar populations.
Producing both the disk and isotropic B-stars population in the same
scenario (and in particular the inner S-stars) requires some fine-tuning.
Specifically, comparing the top heavy mass function of the disk stars,
as suggested by current observations, with the more regular mass function
of the isotropic B-stars disfavors a common origin. One therefore
requires a selective process which works differentially on stars of
different masses. In any case, an additional process would be required
for the migration of stars from the outer to inner regions in the
central pc in order to produce the S-stars close to the MBH.

Two-body relaxation processes work, in principle, differentially on
stars of different masses, through mass segregation (energy equipartition).
For example, relaxation in an isolated stellar disk would somewhat
segregate the more massive stars into a thinner disk \citep{ale+07,per+08b,loc+09},
producing mass stratified populations. However, this effect is relatively
small. Moreover, the most important component for relaxation in the
GC is likely to be the stellar black holes population in the stellar
cusp \citep{per+08b,loc+09,per+09b}, which is dominated by resonant
relaxation \citep{per+09b}. Segregation into two different stellar
populations of different mass regimes is not likely to ensue in this
case. Moreover, such scattering of stars could not produce the population
of S-stars closer to the MBH (\citealp{per+08b}; also Perets et al.,
in prep.). Encounters of binary-single stars \citep{cua+08,per+08c}
can also only have a small effect in producing the isotropic B-stars
population from a thin stellar disk. Perturbations by massive perturbers
\citep{per+07} such as infalling intermediate-mass black holes (IMBHs)
\citep{yuq+07b,gua+09} or other stellar disks \citep[; also Gualandris et al., in
 prep.]{loc+09} do not differentiate between stars of different masses either.

\subsubsection{Efficiency constraints on in-situ formation of the B-stars in the
central pc}

Given the lacking suggestion for a mass differentiating process, a
different route can be taken to explain the different stellar populations.
We can suggest that two distinct epochs of in-situ star-formation
occurred. Two such epochs would still require migration of some of
the stars from the disk to the inner region, where stars do not form
in-situ. One should mention, in this context, that in some cases a
different mass function was found for stars formed in different structures
in the same simulation \citep{hob+09}. We can therefore either suggest
two epochs of star formation happening at different times, or a single
epoch producing two distinct populations. In either case, the rapid
migration producing the S-stars should affect only one of the stellar
populations formed.

Two distinct star formation epochs could naturally produce more massive
and less massive stellar populations. Even if both epochs produced
stellar populations with the same initial mass function, the most
massive stars from the first epoch might already have ended their
life, leaving behind a stellar population of less massive stars. This
scenario could be constrained by observations. We may suggest an interesting
and very general requirement applicable to any scenario in which the
S-stars formed far from their current positions. In any such scenario
only a fraction $f_{mig}$ of the stars formed in some external region
(e.g. the stellar disk, the central pc, or regions outside the central
pc) finally migrate to become S-stars. For a given number of observed
S-stars, $N_{s},$ the parent external population should be $1/f_{mig}$
larger to have \begin{equation}
N_{par}=N_{s}/f_{mig}\label{eq:efficiency}\end{equation}
 stars (compare with similar constraints derived by \citealp{per09}).
Cases in which $f_{mig}$ is small could be strongly constrained by
such requirement. For example, the 'billiard' model for the origin
of the S-stars \citep{ale+04} in which stars from the central pc
are captured close to the MBH through exchanges with SBHs close to
the MBH, is disfavored since the population of similar B-stars in
the central pc is too small to accommodate the required parent population
\citep{pau+06}. Similarly, one can turn to models of a disk origin
for the S-stars, such as the Kozai-like perturbations in the two disks
scenario \citep{loc+09}, the eccentric disk instability model \citep{mad+09},
or the spiral density wave model \citep{gri10}. All of these models,
irrespective of details, suggest that the S-stars formed as part of
a stellar disk, up to a few $0.1$ pc away from the MBH. A small fraction
of the stars formed in the disk migrated through some process to later
on become the currently observed S-stars. The current number of B-stars
inferred in the central $0.5$ pc, but outside the central $0.05$
pc where the S-stars reside, is comparable, and likely somewhat smaller
than the number of S-stars. These models are therefore required to
have a very high migration efficiency in order to explain the origin
of the S-stars. In table \ref{tab:Distribution-of-young-B-stars}
we list the expected number of B-stars in the central $0.05$ pc and
those up to $\sim0.5$ pc, according to the various models. We compare
these numbers with the observationally inferred numbers. We also list
some of the non in-situ formation models, which are discussed later
on. Most of the current formulations of the models are likely excluded,
as can be seen from the table. Whether similar models could still
be consistent with observations under somewhat different conditions
than currently studied, would require additional investigation.

\begin{table*}
\caption{\label{tab:Distribution-of-young-B-stars}}

\begin{tabular}{|l|c|c|c|c|c|}
\hline 
{\footnotesize Model}  & {\footnotesize $f_{mig}$}  & {\footnotesize Observations$^{+}$}  & {\footnotesize Expected $^{\star}$}  & {\footnotesize Observations$^{+}$}  & {\footnotesize Ref.}\tabularnewline
 &  & {\footnotesize{} $<\sim0.05$ pc}  & {\footnotesize {} $<0.5$ pc}  & {\footnotesize {} $<0.5$ pc}  & \tabularnewline
\hline
\hline 
{\footnotesize Massive perturbers (GMC1)}  & {\footnotesize $\sim0.25$}  & {\footnotesize $66\pm15$}  & \textbf{\footnotesize $\sim264$}{\footnotesize{} } & \textbf{\footnotesize $212\pm50$}{\footnotesize {} }  & {\footnotesize This paper}\tabularnewline
\hline 
{\footnotesize External binaries (Empty loss cone)}  & {\footnotesize $\sim0.76$}  & {\footnotesize $66\pm15$}  & \textbf{\footnotesize $\sim87$}{\footnotesize{} } & \textbf{\footnotesize $212\pm50$}{\footnotesize {} }  & {\footnotesize This paper}\tabularnewline
\hline 
{\footnotesize External binaries (Full loss cone)}  & {\footnotesize $\sim0.12$}  & {\footnotesize $66\pm15$}  & \textbf{\footnotesize $\sim539$}{\footnotesize{} } & \textbf{\footnotesize $212\pm50$}{\footnotesize {} }  & {\footnotesize This paper}\tabularnewline
\hline 
{\footnotesize Eccentric disk instability ($e_{init}\ge0.7)$}  & {\footnotesize $\gtrsim0.16-0.32$$^{1}$}  & {\footnotesize $66\pm15$}  & \textbf{\footnotesize $\lesssim206-412$}{\footnotesize{} } & \textbf{\footnotesize $212\pm50$}{\footnotesize {} }  & {\footnotesize \citealp{mad+09}}\tabularnewline
\hline 
{\footnotesize Eccentric disk instability ($e_{init}=0.6)$}  & {\footnotesize $\sim0.07-0.14$$^{1}$}  & {\footnotesize $66\pm15$}  & \textbf{\footnotesize $\sim472-943$}{\footnotesize{} } & \textbf{\footnotesize $212\pm50$}{\footnotesize {} }  & {\footnotesize \citealp{mad+09}}\tabularnewline
\hline 
{\footnotesize Eccentric disk instability$^{2}$ ($e_{init}<0.5)$}  & {\footnotesize $\lesssim0.015-0.03$$^{1}$}  & {\footnotesize $66\pm15$}  & \textbf{\footnotesize $\gtrsim2200-4400$}{\footnotesize{} } & \textbf{\footnotesize $212\pm50$}{\footnotesize {} }  & {\footnotesize \citealp{mad+09}}\tabularnewline
\hline 
{\footnotesize Two disks Kozai$^{3}$}  & {\footnotesize $\lesssim0.045$}  & {\footnotesize $66\pm15$}  & \textbf{\footnotesize $\gtrsim1467$}{\footnotesize{} } & \textbf{\footnotesize $212\pm50$}{\footnotesize {} }  & {\footnotesize \citealp{loc+09}}\tabularnewline
\hline 
{\footnotesize Cluster + IMBH inspiral }  & {\footnotesize $\lesssim0.04$}  & {\footnotesize $66\pm15$}  & \textbf{\footnotesize $\gtrsim1650$}{\footnotesize{} } & \textbf{\footnotesize $212\pm50$}{\footnotesize {} }  & {\footnotesize \citealp{ber+06b,fuj+10}}\tabularnewline
\hline 
{\footnotesize Spiral density wave$^{4}$}  & {\footnotesize $\sim0.01$}  & {\footnotesize $66\pm15$}  & \textbf{\footnotesize $\sim6600$}{\footnotesize{} } & \textbf{\footnotesize $212\pm50$}{\footnotesize {} }  & {\footnotesize \citealp{gri10}}\tabularnewline
\hline 
\multicolumn{6}{|l|}{{\scriptsize $^{+}$Inferred from observations (stars with K magnitude
$13\le14\le16.5$), \citet{bar+10} also H. Bartko, private communication,
2010}}\tabularnewline
\multicolumn{6}{|l|}{{\scriptsize $^{\star}$The expected number normalized to get $\sim53$
stars at $<0.05$ pc}}\tabularnewline
\multicolumn{6}{|l|}{{\scriptsize $^{1}$The range corresponds to a range of total binary
fraction in the disk, taken here to be $0.5-1$.}}\tabularnewline
\multicolumn{6}{|l|}{{\scriptsize $^{2}$Several models studied, most efficient studied
model is shown (in this case a disk with $e_{init}=0.5$) }}\tabularnewline
\multicolumn{6}{|l|}{{\scriptsize $^{3}$Several models studied, most efficient studied
model is shown (4c in \citealp{loc+09})}}\tabularnewline
\multicolumn{6}{|l|}{{\scriptsize $^{4}$The total number of S-stars goes like $(\Delta\alpha/R_{out})^{2}=(0.05/0.5)^{2}\sim0.01$,
see \citealp{gri10}}}\tabularnewline
\hline
\end{tabular}
\end{table*}

\subsection{External star-formation followed by rapid migration}

Another set of scenarios suggest that the GC young stars formed outside
the central pc, where conditions are less hostile to regular star
formation. These scenarios could be used to explain either the origin
of the stellar disk, the origin of the isotropic B-stars, in particular
the S-stars, or both. Note that the extension of S-stars distribution
beyond the central $0.05$ pc is only a recent observational development,
nevertheless, some of the suggested models discuss the possible existence
of such an extended distribution.

\subsubsection{Cluster and IMBH infall}

An infall of a young stellar cluster (with or without an IMBH) into
the central pc was suggested as an alternative scenario for the origin
of the GC young stars \citep{ger01,han+03a,kim+03,por+03,kim+04,lev+05,gur+05,ber+06b,fuj+08,fuj+09}.
Such dissolving cluster is likely to form a stellar disk-like structure,
possibly with additional outlying structures and/or isolated stars
outside the main disk, as observed in the GC. It would also produce
a bias towards more massive stars being concentrated in the central
region of the GC. These massive stars, which were originally segregated
in the innermost regions of the cluster, would be the last to evaporate
from the cluster, i.e. in the central most regions closer to the MBH.
In addition, an IMBH infall has the potential to deplete the inner
regions of the GC, and form a core-like structure \citep{loc+08a,mer09},
as possibly observed \citep{buc+09,do+09,bar+10}. \citet{yuq+07b}
also invoke the existence of an IMBH with mass $>10^{4}\, M_{\odot}$
to help produce higher eccentricities and inclinations for the disk
stars. An infall of an IMBH could also rapidly isotropize the distribution
of stars closest to the MBH, possibly helping explain the currently
observed distribution of the S-stars in the central $0.05$ pc \citep{mer+09,gua+09}.
These properties make this scenario quite attractive. Nevertheless,
the inspiral scenario also has many difficulties, especially in regard
to the origin of the B-stars on which we focus here. We shortly discuss
these in the following.

Infalling clusters may not be able to inspiral in the appropriate
time window for producing the stellar disk. \citet{kim+03} show that
a cluster (without an IMBH) would need to be extremely dense or born
within the central $5$ pc to inspiral fast enough. \citet{kim+04}
find that the presence of an IMBH lessens the requirements on the
cluster density, if it is as massive as 10\% of the cluster mass,
however \citet{gur+04} find that typical masses for IMBHs formed
in simulations of dense clusters (without stellar evolution) are of
the order of $10^{-3}$ of the cluster mass. The formation of appropriate
IMBHs in star clusters may therefore be very difficult. Moreover,
even the formation of lower mass IMBHs through collisions in dense
clusters is found to be difficult when stellar evolution and wind
mass loss are taken into account \citep{yun+08,gle+09,van+09}. The
IMBH cluster infall model is therefore unfavorable given our current
understanding. The mass and orbital separation of a possible IMBH
in the GC are constrained by radio observations of SgrA$^{*}$ \citep[see][for a recent overview of these constraints]{gua+09}.

Another potential issue with the infall model is that an inspiraling
cluster is likely to leave most of its stars behind as it dissolves,
whereas very few young stars are observed outside the central $0.5$
pc. Simulations of an infalling cluster hosting an IMBH \citep{lev+05,ber+06b,fuj+10}
show that the young stars are stripped from the cluster before the
IMBH reaches the central $0.05$ pc. The young stars closest to the
MBH (the S-stars) are therefore not likely to directly originate from
such a scenario. Starting from a cluster of $50$ stars strongly bound
to an inspiraling IMBH (separation of up to $0.06$ pc), \citet{ber+06b}
showed that at most 1-2 stars could reach distances of $0.04$ pc
from the MBH. In other words, $96-98$ percent of the stars were stripped
off when the IMBH was at distances of a few $0.1$ pc from the central
MBH. In terms of the {}``migration efficiency'' defined above, we
get $f_{mig}\le0.04$, where we have replaced the parent population
of the newly formed stars in the disk with the parent population of
the stars originally closely bound to the IMBH. Similar results are
obtained by \citet{fuj+10}, but under somewhat different conditions,
who find one captured S-star (<$0.05$ pc) for $21$ stars stripped
at $<0.4$ pc (M. Fujii, private communication, 2010). For the current
population of S-stars to originate from such a scenario one requires
(Eq. \ref{eq:efficiency}) $N_{p}=N_{s}/f_{mig}\gtrsim1000$ B-stars
to be stripped from the cluster and currently exist in the central
$0.5$ pc; this is much larger than the currently inferred number.

It is also difficult to envision a super-massive cluster containing
$>1000$ massive B-stars within $0.06$ pc of the IMBH for this scenario
to have the necessary apriori conditions.

Even though none of the existing simulations of IMBH inspiral have
been able to produce the S-cluster, including the most recent ones
by \citet{fuj+10} in which massive stars are carried down to distances
no smaller than a few $0.1$ pc from the MBH, we do note that only
\citet{ber+06b} and \citet{fuj+10} have studied in depth the IMBH
infall scenario in the context of the S-stars. Future studies may
explore different conditions than those used there and may bring new
insights on this subject. Similar efficiency arguments as discussed
here could then be applied to test the consistency of such simulations
with current observations.

\subsubsection{Binary disruption}

Another scenario where young stars are formed far from the MBH and
then migrate close to it is the binary disruption scenario. A close
pass of a binary star near a MBH results in an exchange interaction,
in which one star is ejected at high velocity, while its companion
is captured by the MBH and is left bound to it \citep{hil88}. Such
interaction occurs because of the tidal forces exerted by the MBH
on the binary components. A young binary star could therefore be formed
outside the central region and later be scattered onto the MBH on
a highly radial orbit leading to its disruption. Such a scenario was
suggested by \citet{gou+03} to explain the origin of the star S2.
In order for the capture rate of such stars to explain the current
observation of all the GC B-stars, rapid relaxation processes are
required for the binaries to be scattered onto the MBH. Such a model,
suggested by \citet{per+07}, which takes into account scattering
by massive perturbers outside the central 1.5 pc (such as giant molecular
clouds and clumps observed in the GC region and other galactic nuclei;
\citealp{per+07,per+08}) could possibly account for the observed
number of B-stars.

The binary disruption scenario leaves the captured stars on highly
eccentric orbits $(e>0.95)$, and further dynamical evolution is required
in order to explain their currently observed more relaxed eccentricity
distribution. Study of their evolution, which is driven by resonant
relaxation processes \citep{rau+96}, suggest that indeed the more
relaxed, almost thermal eccentricity distribution of the S-stars \citep{per+09b}
could be consistent with their evolution from a much higher initial
eccentricity.

Note that this scenario, like the disk origin models for the S-stars,
discussed above, can be constrained by observations of the parent
population from which the S-stars originate. Current observations
do not exclude this model \citep{per+07,per09}. Future observations
searching for young stars in these regions should, in principle, give
better constraints. Nevertheless, other observational signatures may
be more easily verified or refuted. We discuss these in the following
section which focuses on this scenario.

\section{Evolution of captured stars from binary progenitors outside
the central pc }

In the previous section we discussed various models for the origin
of the B-stars. As discussed above, most of the suggested models can
be strongly constrained and even excluded by simple arguments based
on current observations. In the following we focus on the binary disruption
model, not yet excluded by similar arguments. In a previous study
\citep{per+09b} we focused on the S-stars origin, and used analytic
arguments and $N$-body simulations to show that a binary disruption
origin followed by the (resonant) relaxation of the captured stars
is consistent with their currently observed dynamical properties.
Here we extend this study to explore the dynamics of the isotropic
population of B-stars throughout the central pc. We first describe
in more details the binary disruption scenario. We then discuss the
radial distribution of the captured B-stars according to the model
and how it compares with observations. We also shortly discuss the
mass function of such stars. Finally, we use $N$-body simulations
and analytic arguments to find the expected eccentricity distribution
of the B-stars and its dependence on the distance from the MBH. We
show that the latter distribution is likely to have quite unique properties
and could therefore serve as a strong constraint on the binary disruption
model.

\subsection{The binary disruption scenario: details}

\label{sec:Binary-disruption}

Young stars forming far from the MBH could be part of binaries. If
such a binary is scattered into a highly eccentric orbit, it can pass
the MBH very closely. A close pass of a such binary near the massive
black hole may result in an exchange interaction, in which one star
is ejected at high velocity, while its companion is captured by the
MBH and is left bound to it in a tight orbit. Such interaction occurs
because of the tidal forces exerted by the MBH on the binary components.
Typically, a binary (with mass, $M_{b}$, and semi-major axis, $a_{b}$),
is disrupted when it crosses the tidal radius of the MBH (with mass
$M_{\bullet}$), given by \begin{equation}
r_{t}=\left(\frac{M_{\bullet}}{M_{b}}\right)^{1/3}a_{b}.\label{eq:rtb}\end{equation}
 The capture probability and the semi-major axis distribution of captured
stars were estimated by means of simulations, which show that most
binaries approaching the MBH within the tidal radius $r_{t}(a)$ (Eq.
\ref{eq:rtb}) are actually disrupted \citep{hil91,hil92,bro+06c}.
The harmonic mean semi-major axis for three-body exchanges in equal
mass binaries is \citep{hil91} \begin{equation}
\left\langle a_{1}\right\rangle \simeq0.56\left(\frac{M_{\bullet}}{M_{\mathrm{bin}}}\right)^{2/3}a\simeq\,0.56\left(\frac{M_{\bullet}}{M_{\mathrm{bin}}}\right)^{1/3}r_{t},\label{e:afinal}\end{equation}
 where $a$ is the semi-major axis of the infalling binary and $a_{1}$
that of the resulting MBH-star {}``binary''. Most values of $a_{1}$
fall within a factor $2$ of the mean. This relation maps the semi-major
axis distribution of the infalling binaries to that of the captured
stars: the harder (closer) the binaries, the more tightly bound the
captured stars.

\subsection{Radial distribution of the B-stars}

As discussed above, the initial semi-major axis of the captured star
around the MBH is linearly related to its binary progenitor separation
\citep{hil91}. The binary disruption therefore maps the the distribution
of the binaries separations into the radial distribution of capture
orbits. However, the binary disruption rates could also be dependent
on the binary separation, implying that the cross section for disruption
and hence the size of the loss cone increases with separation. Nevertheless,
the latter is true only for the so-called full loss cone regime (for
a detailed discussion on disruption rates and loss cone regimes see
\citealp{per+07}).

In the empty loss cone regime, the disruption rate $\Gamma$ is independent
of the size of the loss cone, and only depends on the relaxation time
scale of the system \[
\frac{d\Gamma}{da}\propto\frac{dN_{bin}(a)/da}{t_{r}}\]
 (following Eq. 17 in \citealp{per+07}), where $N_{bin}(a)$ is the
number of binaries with separation $a$ and $t_{r}$ is the relaxation
time of the system. Since the separation of the captured star is linearly
related to the progenitor stellar binary separation, the radial distribution
of captured stars follows the binary separation distribution.

In the case of the full loss cone regime,\[
\frac{d\Gamma}{da}\propto\frac{dN_{bin}(a)}{da}q\propto\frac{dN_{bin}(a)}{da}a,\]
 (following Eq. 18 in \citealp{per+07}), where $q$ is the the tidal
disruption radius of a given binary, which is linearly related to
the binary separation (see Eq. \ref{eq:rtb} above). In this case
the radial distribution of the captured stars does not directly follow
the binary semi-major axis distribution.

We therefore find that the two extreme cases of empty and full loss
cone regimes produce a different disruption rate dependence on the
binary separation and therefore imply different radial distributions
for the captured stars.

The distribution of semi-major axes of massive binaries in the Solar-neighborhood
follows a log-constant distribution. In the empty loss cone regime
the disruption rate is independent of the binary separations. Therefore,
the radial distribution of captured stars in the GC should directly
correspond to the binary separation distribution, according to Eq.
\ref{e:afinal}. We therefore expect a log-constant, or a cumulative
distribution of $N(<r)\propto\log r$ for the captured B-stars, under
the assumption that the GC binaries have similar properties to those
in the Solar neighborhood. In the full loss cone regime, the binary
disruption rate is linearly dependent on the binary separation, and
wider binaries have a higher chance of being disrupted compared to
shorter period binaries. We therefore expect a cumulative distribution
which goes like $N(<r)\propto r.$

\begin{figure}[!]
 \includegraphics[scale=0.35]{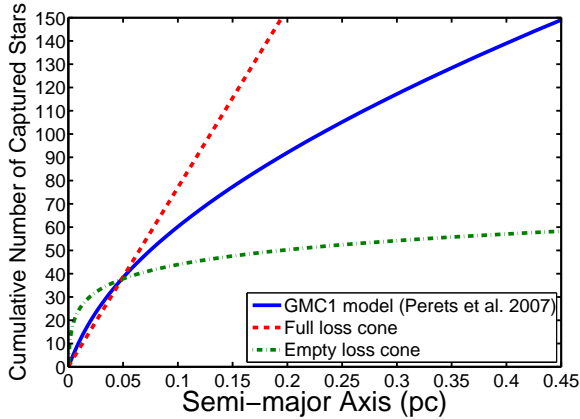} \caption{The cumulative distribution of semi-major axes for captured stars
up to $0.5$ pc. Several models are shown, including an empty loss
cone, a massive perturbers refilled loss cone \citep{per+07}, and
a full loss cone (see text). The empty and full loss cone models are
normalized normalized to have $\sim40$ B-stars within $0.05$ pc.
The massive perturbers model is not normalized, and uses the GMC1
mode \citep{per+07} assumptions, but with revised binary separtions
distribution (see text). }

\label{f:radial-dist} 
\end{figure}

In fig. \ref{f:radial-dist} we show the cumulative distribution of
semi-major axes for captured stars up to $0.5$ pc, normalized to
have $\sim40$ B-stars within $0.05$ pc, fairly consistent with observations
(which infer $66\pm15$ stars; \citet{bar+10}; also H. Bartko, private
communication, 2010). Also shown in fig. \ref{f:radial-dist} is the
expected distribution of captured B-stars in the scenario of binary
disruption induced by massive perturbers \citep{per+07}, which produces
$\sim40$ B-stars up to $0.05$ pc. The distributions are obtained
using the detailed calculations described in \citet{per+07}, which
take into account the relevant relaxation processes. The GMC1 model
described there is assumed for the massive perturber scenario. A log-constant
binary semi-major axis distribution is assumed for the progenitor
binary population (compare with the log-normal distribution assumed
in \citealt{per+07}, based on the data available at the time; \citet{kob+07},
which was later revised to be a log-constant distribution). This latter
scenario produces a distribution intermediate between the extreme
empty loss cone regime and the extreme full loss cone regime.
Interestingly, the captured stars radial distribution could therefore 
give a handle on the type of relaxation processes in the GC, given the
assumptions on the binaries distribution. 
According to GMC1 model, the expected number of B-stars in the central
$0.5$ pc is $\sim260$ stars (when noramlized to 66 at $<0.05$ pc).
 The empty and full loss cone regimes,
when normalized to $66$ captured stars at $0.05$ pc, predict, respectively,
$\sim90$ and $\sim540$ stars captured up to $0.5$ pc. In order
to get a simple comparison with other models for the B-stars origin,
we can define an effective {}``migration efficiency'' for this model.
In this case we replace the parent population with the population
of stars captured up to $0.5$ from the MBH, i.e. strictly speaking,
this is not the parent population but the stellar population which
one would like to probe when comparing the models studied here. We
find in this case $f_{mig}\sim0.25$ and $\sim0.12$ for the empty
and full loss cone regimes, respectively (see also table \ref{tab:Distribution-of-young-B-stars}).

\subsection{Eccentricity distribution of the B-stars}

The periapse of the captured star in the binary disruption scenario
is at $r_{t}$, and therefore its eccentricity is very high \citep{hil91,mil+05},
\begin{equation}
e\!=\!1-r_{t}/a_{1}\!\simeq1-1.8(M_{\mathrm{bin}}/M_{\bullet})^{1/3}\!\gtrsim\!0.97\label{eq:capture-eccentricity}\end{equation}
 for values typical of binaries and the MBH in the GC ($M_{bin}=20\, M_{\odot}$;
$M_{\bullet}=4\times10^{6}\, M_{\odot}$).

For comparison, the initial eccentricity distribution of stars formed
in a stellar disk is likely to be biased to much more circular orbits.
For example, the currently observed eccentricity distribution of the
young O and WR stars observed in the clockwise disk peaks at $e\sim0.35$.

It is interesting to note, however, that the resonance relaxation
timescales in the GC increase with distance from the MBH \citep{hop+06a}.
Therefore, stars captured/formed further away from the MBH are likely
to have a less relaxed eccentricity distribution, i.e. over time we
may expect a correlated eccentricity-distance distribution to be produced,
with stars more distant from the MBH having an eccentricity distribution
which much more closely resembles the original one. Far from the MBH,
where the resonant relaxation timescales are much larger than the
lifetimes of the B-stars (see e.g. \citealp{hop+06a}), stars should
keep their original eccentricity distribution, i.e. highly eccentric
orbits for captured stars, and likely low eccentricity orbits for
stars formed in a stellar disk. Stars captured a few $0.1$ pc away
from the MBH, for example, are likely to still conserve their initial
eccentricities. Closer to the MBH captured stars could have a relaxed
(thermal eccentricity) distribution even after short times, although
the captured stars population relaxes to a thermal eccentricity distribution
much faster than a stellar population with initially low typical eccentricity
\citep{per+09b}).

\subsubsection{Results of N-body simulations }

In order to study the eccentricity distribution of the captured stars
and its radial dependence more quantitatively we performed detailed
$N$-body simulations. We used the $\phi\, GRAPE$ direct summation
software running on \texttt{gravitySimulator}, the 32-node cluster
at the Rochester Institute of Technology that incorporates GRAPE accelerator
boards in each of the nodes \citep{har+07}. We assume that the main
contribution to the dynamical relaxation of stars in the GC comes
from stellar black holes (SBHs). We consider a distribution of SBHs
following the results of Fokker-Plank calculations of the distribution
of $10\, M_{\odot}$ SBHs around a MBH found by \citep{hop+06b}.
We use 16000 SBHs of $10\, M_{\odot}$ with an isotropic radial, $r^{-2}$,
distribution between $0.04$ pc and $0.8$ pc around a MBH of $3.6\times10^{6}\, M_{\odot}$
(the simulation of stars closer to the MBH requires much longer simulation
time; we discussed the evolution of such stars in \citealp{per+09b}).
In order not to be affected by boundary effects we focus on stars
in the range $0.08--0.5$ pc from the MBH. In all of our simulations
we force the MBH to be at rest at the origin of the system since the
Brownian motion is expected to be quite low \citep{mer05}. Note
that the full cusp of low mass stars was not included in our simulations
(this is beyond our current computatational capabilites). Such stars 
could somewhat accelarate the resonant relaxation processes as
more stars contribute to the process. However, the larger mass
contributing to the cusp potential would also
produce precession of the stellar orbits, thereby producing enhanced 
quenching of resonant resonant relaxation.  
 The neglection of the full cusp potential therefore suggests
that our N-body results likely give an upper limit to the relaxation
rate due to resonant relaxation. The signature of the initial capture conditions might therefore even better preserved than suggested by the current simulations.  
As in our previous simulations \citep{per+09b}
general relativisitic precession due to the MBH is not included. The
latter component, however, is not important for stars at the distances
from the MBH discussed here.

The {}``captured stars'' followed by our simulations are those stars
with initial eccentricities of $0.95\le e\le0.99$. %
\begin{figure}[!]
\includegraphics[scale=0.35]{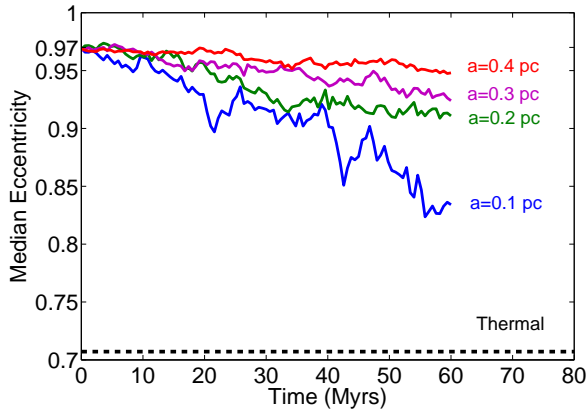} \caption{The temporal evolution of the median eccentricity of the captured
B-stars in the central $0.5$ pc, at various distances from the MBH. }

\label{f:eccs-temporal} 
\end{figure}

In fig. \ref{f:eccs-temporal} we show the temporal evolution of the
median eccentricity of the captured B-stars in the central $0.5$
pc, at various distances from the MBH. The MS lifetimes of $7-15\, M_{\odot}$
stars are $\sim10-50$ Myrs. 
in the future.

\begin{figure*}[!]
 \includegraphics[scale=0.25]{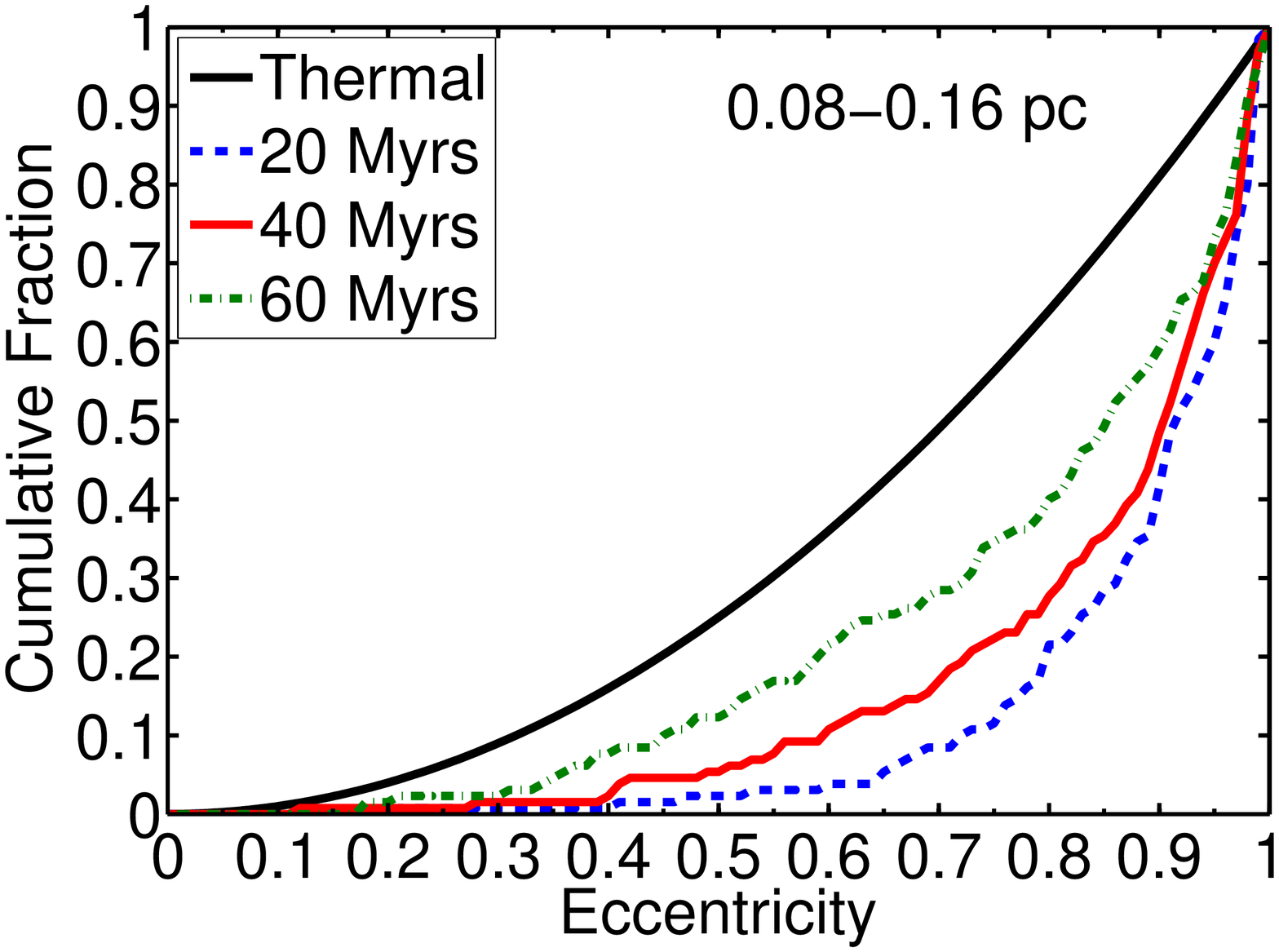} \includegraphics[scale=0.25]{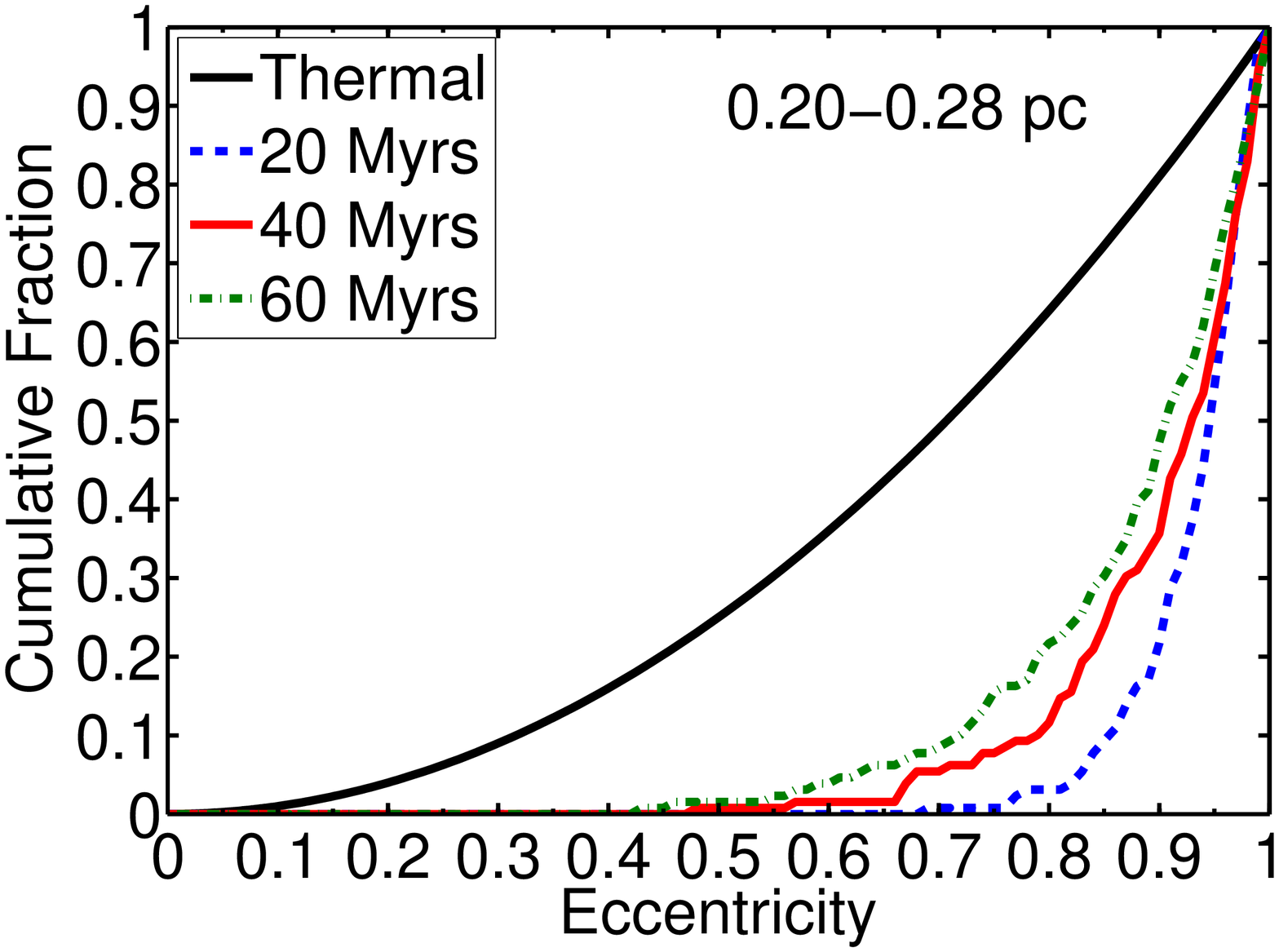}
\includegraphics[scale=0.25]{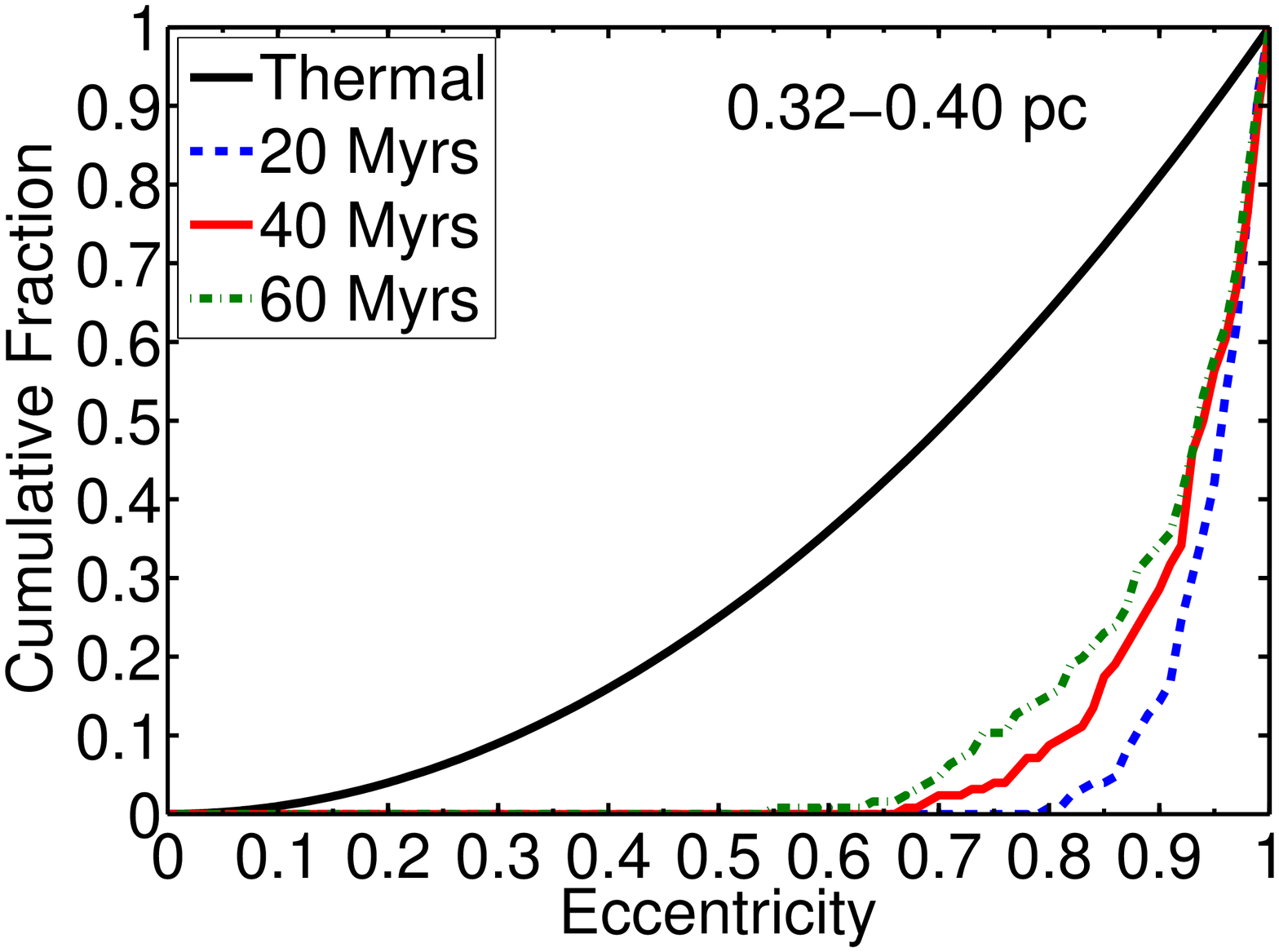} \caption{The cumulative eccentricity distribution in three different distance
regimes, after $20$, $40$ and $60$ Myrs of evolution.  Also shown for comparison is a thermal eccentricity distribution
($f(e)=2e$ ; cumulative $cf(e)=e^{2}$). Captured stars have initial
eccentricities between 0.95 and 0.99.}

\label{f:eccs-cumulative} 
\end{figure*}

In figs. \ref{f:eccs-cumulative}a-\ref{f:eccs-cumulative}c we show
the cumulative eccentricity distribution in three different distance
regimes and at $20$, $40$ and $60$ Myrs of evolution. Also shown
for comparison is a thermal eccentricity distribution ($f(e)=2e$
; cumulative $cf(e)=e^{2}$).

\begin{figure}[!]
 \includegraphics[scale=0.35]{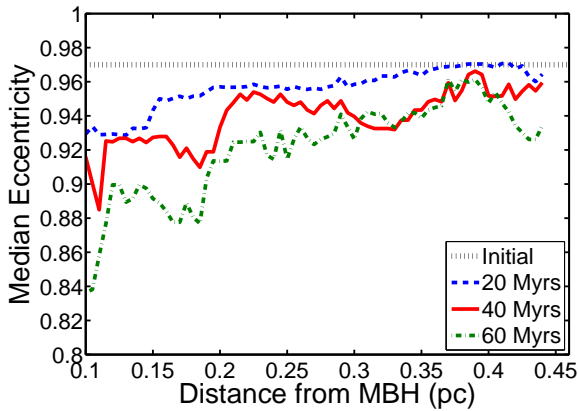} \caption{The median eccentricity of
the captured B-stars vs. their separation from the MBH after $20$, $40$ and $60$ Myrs of evolution. For comparison
the median eccentricity of a relaxed thermal population is $\sim0.7\,(\sqrt(1/2)$.  Captured stars have initial median eccentricity
of $\sim0.97$ (top dotted line). }

\label{f:eccs-distance} 
\end{figure}

In fig. \ref{f:eccs-distance} we show the median eccentricity of
the captured B-stars vs. their separation from the MBH at $20$, $40$
and $60$ Myrs of evolution. As can be seen in this figure, the binary
capture scenario provides a qualitatively unique signature, in which
the typical eccentricity of the stars rises with distance. The exact
relaxation rate in the GC is dependent on the number density of stars
and SBHs in this environment, which is still poorly constrained. For
our assumptions on the number density of SBHs, captured stars a few
$0.1$ pc away from the MBH have very high $(>0.9)$ eccentricities,
whereas stars are more relaxed at closer to the MBH,  as also found in previous
studies on the captured population even closer to the MBH \citep{per+09b}.
Faster relaxation rates (e.g produced by a higher density of SBHs
and/or the additional contribution to relaxation process by low mass
stars, not included here) would still produce the qualitative signature
of a typical eccentricity rising with distance, but may quantitatively
show lower eccentricities, and a relaxed population up to larger distances
from the MBH. Conversely, a stellar population with initially low
($<0.5$) eccentricities, as might be expected from a stellar disk
population, may show the opposite picture, with progressively lower
eccentricities at larger distances from the MBH. For this reason the
distribution of eccentricities vs. distance could be a key observational
property for discriminating models for the origin of the young B-stars.
We do note, however, that very close to the MBH (the innere regions of the S-cluster art a few 0.1 pcs from the MBH), where general relativistic precession due to the MBH becomes important. At these regions resonant relaxation might be quenched, and capture stars could again show higher eccentricities.

\subsection{Mass function of the captured stars}

The mass function (MF) of captured stars is likely to be regular at
the high mass regime ($\gtrsim3\, M_{\odot}$), i.e. reflecting the
mass function of stars far from the MBH, where regular star-formation
can take place, and where the progenitor binaries form. The currently
observed MF, however, may slightly differ from the regular MF due
to several reasons.

(a) Binary fractions and the distribution of binary orbital properties
are mass dependent. Massive (OB) stars have higher binary fractions
than lower mass stars. Massive binaries are also typically much more
compact (a log constant distribution of periods vs. a log-normal distribution
for low mass stars; e.g. \citealp{duq+91,kob+07}). The binary disruption
process requires binary progenitors for the captured stars. It also
maps the binary period into the capture separation from the MBH as
discussed above. For both these reasons, a larger fraction of massive
stars is likely to be captured in this scenario, compared with the
fraction of low mass stars binary disruption (relative fractions;
in terms of absolute frequency of low mass binary stars, they are
still more frequent than high mass binary stars). In addition, massive
stars are typically captured much closer to the MBH than low mass
stars. Therefore, the population of captured stars in the GC, especially
in the central pc, tends to be biased towards massive OB stars.

(b) Longer living stars (hence lower mass stars) captured at earlier
times may have a higher probability of being disrupted by the MBH
during their dynamical evolution \citep{per+09b}. On even longer
timescales (two-body relaxation timescales), they can migrate from
their original capture orbit. For this reason, the observed population
of captured stars is biased towards stars which were recently captured,
i.e. selected against long living stars captured at earlier times.

(c) A small contribution from Kozai induced merger of captured binary
stars, induced by Kozai resonances near the MBH, may contribute a
small fraction of more massive stars \citep{ant+09,per09b,per+09c}.

Currently, the MF of the probed B-stars population is consistent with
a continuous star formation (with regular IMF) over the last $\sim50$
Myrs \citep{bar+10}. The range of probed masses is currently quite
limited, and low mass stars are not included in the analysis (and
therefore point (a) is not directly verified, given our limited knowledge
on the binary distribution of OB stars of different masses. For this
limited range, the observed MF is qualitatively consistent with the
expected one. Future observations probing fainter stars will provide
much better data and therefore stronger constraints on the mass function.

\section{Discussion}

Any model suggesting a common origin for all/most of the B-stars in
the GC is required to produce both the stars in the inner region $(<0.05$
pc) and those in the outer region ($<0.5$ pc or more). Most of our
discussion on the constraints on different models for the origin of
these stars can therefore be summarized by two simple qualitative
properties: (i) the ratio between the number of stars in the inner
region and the outer region of the GC, (ii) the radial eccentricity
distribution trend in these regions.

Table \ref{tab:Distribution-of-young-B-stars} summarizes the predictions
of the various theoretical models for the number of B-stars in the
inner/outer regions, compared to current observations. As can be seen,
most models grossly over-predict the number of B-stars in the outer
region, up to $0.5$ pc, given the observed number of B-stars in the
central $0.05$ pc. Such constraints could easily be studied and extended/generalized
for any new model suggested for the B-stars origin.

Currently, we find that the most consistent explanation for the complete
population of B-stars in the GC is an external formation followed
by binary disruption (likely driven by massive perturbers). The eccentricity
distribution of the captured stars in this scenario should be dependent
on the distance from the MBH. Higher eccentricities should be observed
for stars captured at larger distances.

The formation of binary B-stars in a very eccentric stellar disk with
$e_{init}>0.7$ could also be consistent with the ratio of inner to
outer B-stars, for a high enough binary fraction in the stellar disk.
The radial distribution of the stars in this case is likely to differ
from that of the massive perturbers scenario, as the population of
B-stars would include both the progenitor population formed in the
stellar disk and the captured stars. These two populations are likely
to have different radial distributions, and their combined distribution
is likely to show a bi-modality, with one peak at the inner region
close to the MBH, dominated by the captured B-stars, and an additional
peak corresponding to the peak in the stellar disk. From the simulations
by \citet{mad+09}, it seems that the binary disruption is only weakly
dependent on the binary separation. This is reminiscent of the empty
loss cone regime discussed above. The radial distribution of the captured
stars should therefore correspond to a direct mapping of the binaries
separations distribution in this case. The eccentricity distribution
of the B-stars should also be a combination of two populations. A
more detailed study of this scenario, however, is beyond the scope
of this paper and will be studied elsewhere (see \citealp{mad+09}
for a discussion of the basic scenario).

\section{Summary}

In this study we explored the dynamical constraints on the origin
of the young B-stars in the Galactic center. We showed that most of
the currently studied models could not consistently explain both the
population of the S-stars (the B-stars at $<0.05$ pc from the MBH),
and the extended population of young B-stars up to $0.5$ pc. In these
models the number of B-stars in the inner and outer regions are incompatible.

We used analytical arguments and N-body simulations to study the capture
and the dynamics of the B-stars in the external formation+binary disruption
scenario, which is one of the models consistent with the observed
B-stars numbers at different regions of the GC. Using our analysis
we provide detailed predictions for the distribution of the B-stars
dynamical properties, which could serve as observational signatures
to further support, constrain or refute this model.

\acknowledgements{We would like to thank Tal Alexander for helpful discussions as well
as Hendrik Bartko and Michiko Fujii for providing us with data from
their observations and simulations, respectively. We would also like
to thank David Merritt for the use of the GRAPE cluster at the Rochester
Institute of Technology. HBP would like to thank the Israeli Science
Foundation, the Bi-national Fulbright program and the Israeli industrial
and commercial club for their support through the BIKURA (FIRST) and
Ilan Ramon-Fulbright fellowships. AG is supported by grant NNX07AH15G from NASA.}

\bibliographystyle{apj}


\end{document}